\documentclass[preprint,endfloats,prb,superscriptaddress,showpacs]{revtex4}

\usepackage{amsmath,graphicx,natbib}

\newcommand{\unial}{U$_3$Ni$_5$Al$_{19}$}
\newcommand{\gdnial}{Gd$_3$Ni$_5$Al$_{19}$}
\newcommand{\thnial}{Th$_3$Ni$_5$Al$_{19}$}
\newcommand{\ceni}{Ce$_7$Ni$_3$}

\begin{document}


\title{High pressure investigation of the heavy-fermion antiferromagnet U$_3$Ni$_5$Al$_{19}$}

\author{E. D. Bauer}
\affiliation{Los Alamos National Laboratory, Los Alamos, NM 
87545, USA}

\author{V. A. Sidorov} \altaffiliation{Institute for High Pressure Physics, Russian Academy of Sciences, 142190 
Troitsk,  Russia}
\affiliation{Los Alamos National Laboratory, Los Alamos, NM 
87545, USA}

\author{S. Bobev} \altaffiliation{Present address:  Department of Chemistry, University of Delaware, Newark, Delaware 19716}
\affiliation{Los Alamos National Laboratory, Los Alamos, NM 
87545, USA}

\author{D. J. Mixson}
\affiliation{Department of Physics, University of Florida, Gainesville, FL 32611, USA}

\author{J. D. Thompson}
\affiliation{Los Alamos National Laboratory, Los Alamos, NM 
87545, USA}

\author{J. L. Sarrao}
\affiliation{Los Alamos National Laboratory, Los Alamos, NM 
87545, USA}

\author{M. F. Hundley}
\affiliation{Los Alamos National Laboratory, Los Alamos, NM 
87545, USA}


\date{\today}


\begin{abstract}

Measurements of magnetic susceptibility, specific heat, and electrical resistivity at applied pressures up to 55 kbar have been carried out on single crystals of the  heavy-fermion antiferromagnet \unial, which crystallizes in the \gdnial{} orthorhombic structure with two inequivalent U sites.  At ambient pressure, a logarithmic $T$-dependence of the specific heat and $T-$linear electrical resistivity below 5 K indicates non-Fermi liquid (NFL) behavior  in the presence of  bulk antiferromagnetic order at $T_N=23$ K.  Electrical resistivity measurements reveal a crossover from non-Fermi liquid  to Fermi liquid behavior at intermediate pressures between 46 kbar and 51 kbar, followed by a return to NFL $T^{3/2}$ behavior at higher pressures.   These results provide evidence for an ambient pressure quantum critical point and an additional antiferromagnetic instability at $P_c \approx60$ kbar.

\end{abstract}


\maketitle

A considerable amount of experimental and theoretical effort has been devoted  in recent years to the investigation of quantum criticality in $f$-electron heavy-fermion materials (for a recent review, see Ref. \onlinecite{Stewart01}).  Until now, most research has focussed on the behavior of Ce-based heavy-fermion systems in the vicinity of an antiferromagnetic (AFM) quantum critical point (QCP) in which the N{\'{e}}el temperature $T_N$ is tuned to absolute zero by an external control parameter such as composition $x$, pressure $P$, or magnetic field $H$.  Measurements under applied pressure on these systems have been particularly useful in accessing the QCP and also exploring the unusual power law or logarithmic $T$-dependences of the physical properties, characteristic of non-Fermi liquid (NFL) behavior, found near the critical point.  For instance, at a critical pressure $P_c=28$ kbar necessary to completely suppress antiferromagnetism in CePd$_2$Si$_2$,\cite{Mathur98} the electrical resistivity exhibits a power law, i.e., $\rho - \rho_0 = AT^n$, with $n=1.2$ over an extended range in temperature, in contrast to the $T^2$ Fermi-liquid behavior expected for a simple metal.  The antiferromagnetic Ce-based heavy-fermion compounds have proved unstable to the formation of superconductivity in the vicinity of the AFM QCP.  The transition temperatures are often quite low, $T_c \sim 0.4$ K (0.2 K) for CePd$_2$Si$_2$  (CeIn$_3$) at $P=28$ kbar (26 kbar);\cite{Mathur98} more recently, superconductivity has been observed at $T_c \sim 2$ K in antiferromagnetic CeRhIn$_5$ above 15 kbar,\cite{Hegger00} more than half the value of the N{\'{e}}el temperature at ambient pressure ($T_N=3.8$ K).  The occurrence of superconductivity near the AFM QCP where spin fluctuations are strongest is indicative of an unconventional magnetically mediated  pairing mechanism in these compounds.  Therefore, it is of interest to investigate pressure-induced quantum criticality in antiferromagnetic U-based heavy-fermion materials as comparatively fewer such studies have been performed.\cite{Thompson94}  To this end, we present measurements of magnetic susceptibility, specific heat, as well as electrical resistivity up to 55 kbar of the heavy-fermion antiferromagnet \unial.

\unial{} crystallizes in the orthorhombic Gd$_3$Ni$_5$Al$_{19}$ structure\cite{Gladyshevskii92a} (space group Cmcm, No. 63) with two inequivalent U sites (one U atom in $4c$ and two U atoms in $8f$) which can be thought of as an intergrowth between the imaginary structures of YbNiAl$_4$ (with the orthorhombic YNiAl$_4$-type structure \cite{Rykhal72}) and that of Yb$_2$Ni$_4$Al$_{15}$ (monoclinic).  A recent report\cite{Haga04} concluded that \unial{} orders antiferromagnetically at $T_N=23$ K, due to a prominent feature in $\chi(T)$ for $H||c$ involving one of the two distinct U sites, while the other site showed no sign of magnetic order down to 50 mK.  

\begin{figure}[htbp]
\includegraphics[width=5.0in]{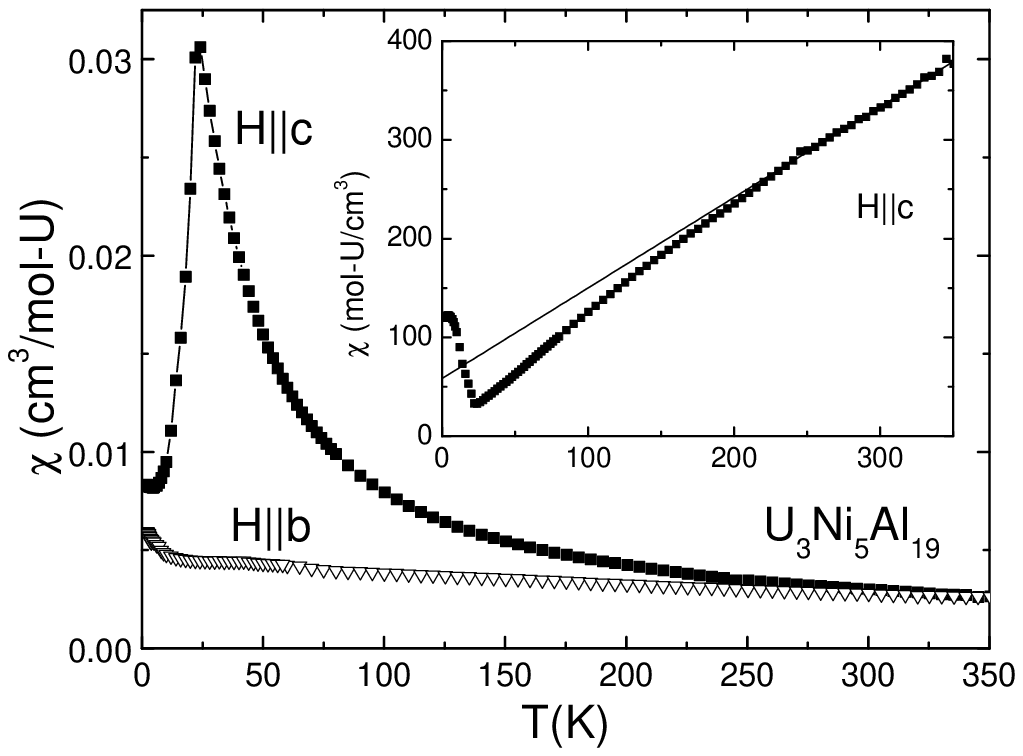}
 \caption{Magnetic susceptibility $\chi(T)$ in a magnetic field $H=0.1$ T along the $b$- and $c$-axes.  Inset: Inverse magnetic susceptibility $\chi^{-1}(T)$ for $H||c$.  The solid line is a linear fit of the data.}
\label{chi}
\end{figure}
 Single crystals of \unial{} were grown in Al flux.  The elements were placed in the ratio U:Ni:Al=1:1:10 in an alumina crucible sealed under vacuum in a quartz tube.  The sample was heated to 1100 $^{\circ}$C and kept at that temperature  for 4 hr., then slowly cooled to 650 $^{\circ}$C at 7 $^{\circ}$C hr$^{-1}$, at which point excess Al flux was removed in a centrifuge.  The resulting crystals were needles with typical dimensions 1$\times$1$\times$5mm$^3$.  The orthorhombic Gd$_3$Ni$_5$Al$_{19}$ structure was confirmed by single crystal x-ray diffraction with lattice parameters $a=4.0850(2)$ \AA, $b=15.9305(8)$ \AA, and $c=26.959(1)$ \AA{} (further details of the structural refinement can be found in Ref. \onlinecite{CSD_u3ni5al19}).  Magnetic susceptibility measurements were performed in a commercial magnetometer from 2-350 K in a magnetic field $H=0.1$ T both parallel and perpendicular to the long axis of a single crystal. It was then concluded on the basis of other measurements\cite{Haga04} that the long axis of the crystal was the $b$-axis.   Specific heat measurements were made using a commercial calorimeter from 0.4 K to 50 K on a collection of single crystals using an adiabatic method.  Electrical resistivity measurements under pressure were carried out using a  profiled
toroidal anvil clamped device with
 anvils supplied with a boron-epoxy gasket and Teflon capsule, containing pressure-transmitting liquid, sample and a pressure sensor.\cite{Khvostantsev98}  The pressure was determined from the variation of the superconducting transition of lead using the pressure scale of Eiling and Schilling.\cite{Eiling81}   A standard four-probe technique was performed using an LR-700 Linear Research bridge operating at a current of 1 mA applied along the $b$-axis  of the crystal.

The magnetic susceptibility $\chi \equiv M/H$ of \unial{} in a magnetic field $H=0.1$ T along various crystallographic directions is shown  Fig. \ref{chi}.  A clear signature of a magnetic transition, presumably antiferromagnetic, is found in $\chi_c$ at $T_N= 23$ K, while $\chi_b$ exhibits temperature independent paramagnetism, indicating marked magnetic anisotropy.  The $\chi_c$ data  can be fit to a Curie-Weiss law above 250 K  yielding an effective moment $\mu_{eff}=3.0$ $\mu_B$/U atom and Curie-Weiss temperature $\theta_{CW}=-79$ K as shown in the inset of Fig. \ref{chi}.

\begin{figure}[htbp]
\includegraphics[width=5.0in]{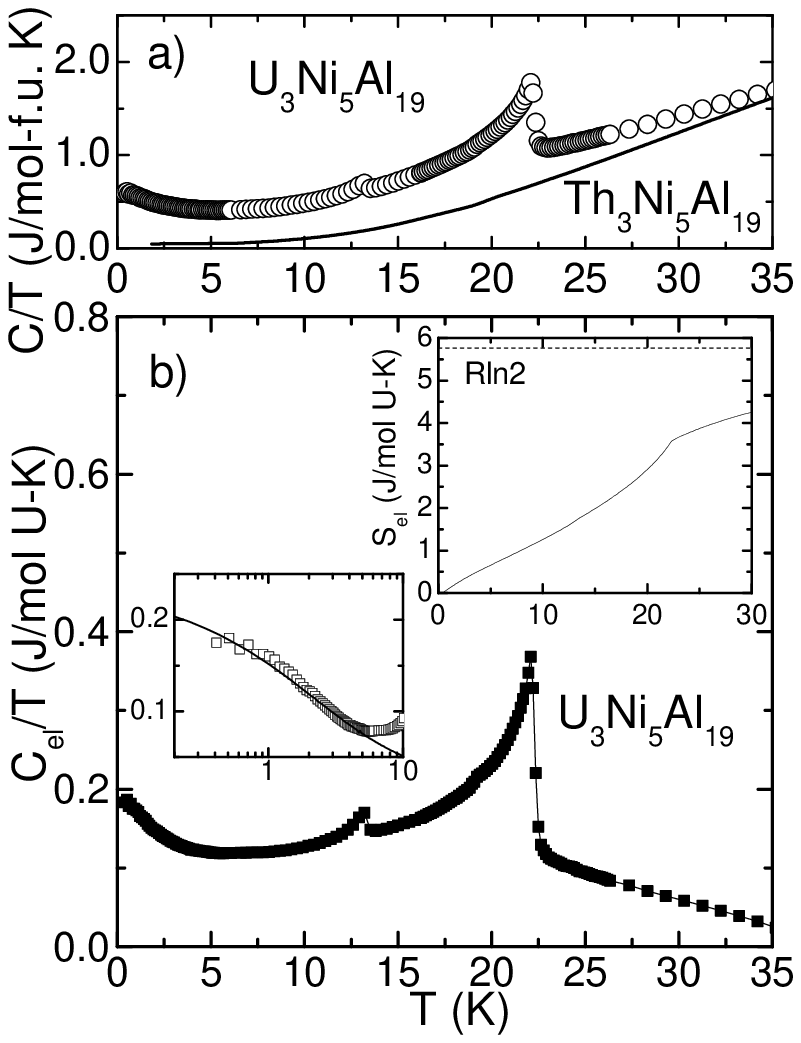}
 \caption{a) Specific heat $C(T)$ of \unial{} and nonmagnetic \thnial.  b) Electronic contribution $C_{el}$ to the specific heat of \unial, plotted as $C_{el}/T$ vs $T$.  Lower inset: $C_{el}(T)/T$ below 10 K.  The solid line is a fit to the spin-fluctuation theory discussed in the text. Upper inset: Electronic entropy $S_{el}(T)$.}
\label{cp}
\end{figure}
The specific heat $C(T)$ of \unial{} and nonmagnetic \thnial{} is shown in Fig. \ref{cp}a.  The specific heat of \thnial{} is characterized by a Sommerfeld coefficient $\gamma=1$ mJ/mol-Th K$^2$ and a Debye temperature $\theta_D=370$ K.  An anomaly in \unial{} is observed at $T_N=23$ K confirming bulk AFM order.  A second, much smaller feature  is found at $\sim 13$ K and is attributed to an impurity phase as no such feature is observed in $\chi(T)$ or $\rho(T)$, although a spin-reorientation of the U moments cannot be ruled out.  Below 5 K, $C_{el}/T$ exhibits a non-Fermi liquid $-log(T)$ dependence as the temperature is lowered characteristic of a system in proximity to a quantum critical point. 
The electronic contribution $C_{el}$ to the specific heat of \unial, plotted as $C_{el}/T$ vs $T$, is shown  in Fig. \ref{cp}b, where the nonmagnetic contribution of \thnial{} has been subtracted from the $C(T)$ data.    A significant quasiparticle mass enhancement is indicated by the large value of $C_{el}/T$ =   185 mJ/mol-U K$^2$ at $T=0.4$ K. The electronic entropy $S_{el}(T) = \int (C_{el}/T) dT$ released below $T_N$ amounts to $S_{el}$(23 K)=$3.9$ J/mol-U K as shown in Fig. \ref{cp}b.   This value amounts to (0.67)R$ln(2)$, considerably less than  R$ln(9)$ or R$ln(10)$ expected for a $5f^2$ ($J=4$) or $5f^3$ ($J=9/2$) electronic configuration, respectively.  
  The $C_{el}/T$ data below 5 K can be fit by the spin-fluctuation model of Moriya and Takimoto\cite{Moriya95} describing the contribution of critical spin fluctuations to the specific heat with input parameters:  the distance from the QCP, $y_0$, characteristic energy $T_0$ proportional to the exchange energy, and Sommerfeld coefficient for non-critical Fermions $\gamma_0$.  As displayed in Fig. \ref{cp}b, a fit of the data to this theory yields  $y_0=0.001$, $T_0=3.2$ K, and $\gamma_0=80$ mJ/mol-U K$^2$, suggesting \unial{} is close to a $P=0$ quantum critical point.  
After subtracting this NFL contribution from the $C_{el}(T)$ data of \unial, the magnetic contribution to the specific heat $C_{mag}$ (not shown) is  reasonably well-described by a $T^3$ power law in the antiferromagnetic state characteristic of antiferromagnetic spin-wave excitations.\cite{Ashcroft76}  
\begin{figure}[htbp]
\includegraphics[width=5.0in]{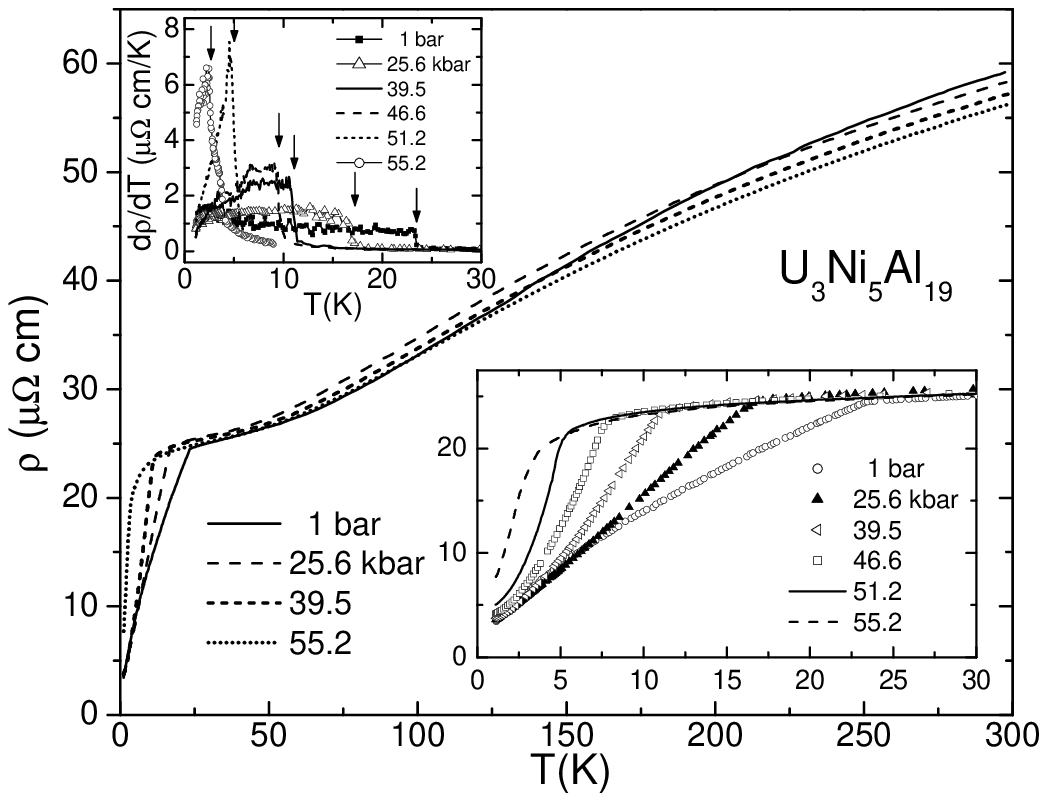}  
 \caption{Electrical resistivity $\rho(T)$ of \unial{} at various pressures $P$.  Lower inset:  $\rho(P,T)$ below 30 K.  Upper inset:  $d\rho/dT$ vs $T$ at various pressures.  The arrows indicate the  N{\'{e}}el temperature $T_N$.}
\label{rho}
\end{figure}

The results of electrical resistivity measurements under pressure of \unial{} are presented in Fig. \ref{rho} at four different pressures.  The $\rho(P,T)$ curves have ``s-shaped" curvature typical of spin fluctuations systems.  The magnetic phase transition is readily visible as a kink in $\rho(T)$ at $T_N=23$ K at ambient pressure.  Upon further cooling, the resistivity is linear in temperature below 5 K consistent with the NFL behavior observed in specific heat.  The application of pressure suppresses the N{\'{e}}el temperature to $T_N=2.4$ K at $P=55.2$ kbar, the highest pressure reached in this experiment, as shown in the lower inset of Fig. \ref{rho}.  The derivative $d\rho/dT$ is characteristic of a second order phase transition \cite{Fisher68} and reveals a similar suppression of the magnetic transition as displayed in the upper inset of Fig. \ref{rho}.

\begin{figure}[htbp]
\includegraphics[width=6.0in]{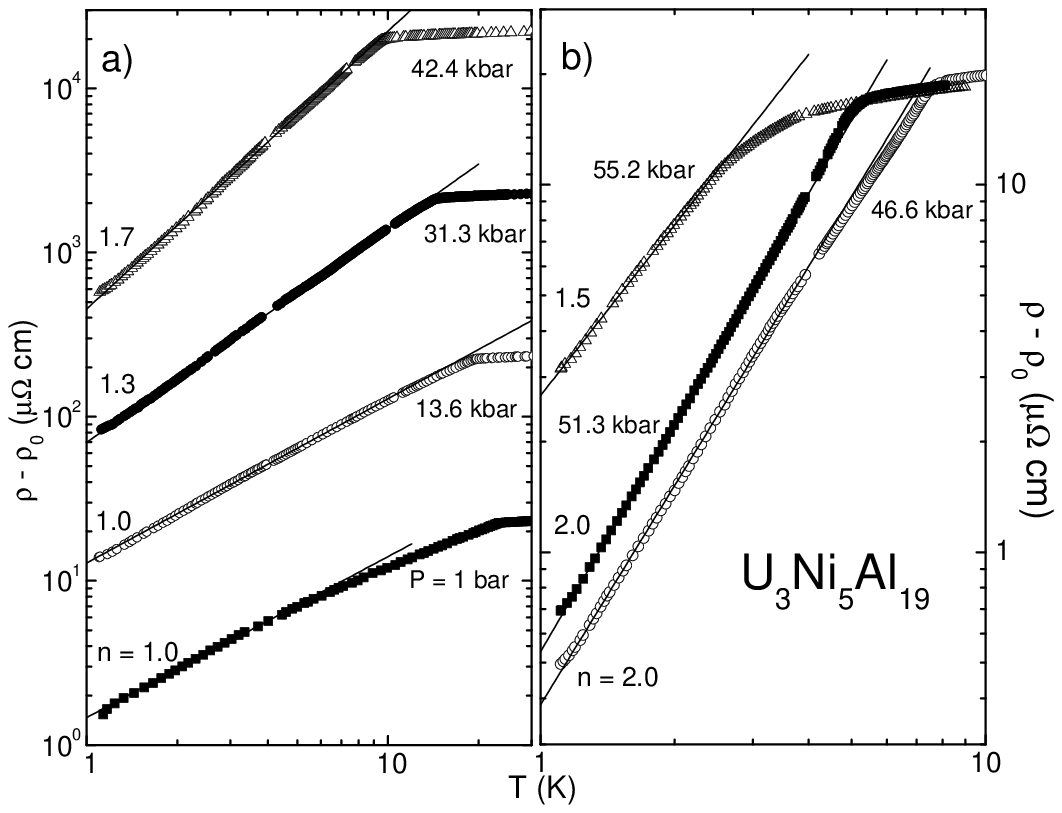}
 \caption{a) $\rho - \rho_0$ vs $T$ for $P \leq 42.4$ kbar on a log-log scale.  Each of the curves have been shifted vertically by one decade from the curve below it for clarity. b) $\rho - \rho_0$ vs $T$ for 46.6 kbar $\leq P \leq 55.2$ kbar (the curves have not been shifted vertically).  The solid lines in a) and b) are power law fits of the data  to $\rho - \rho_0 = AT^n$, where the exponent $n$ is indicated next to each curve.   }
\label{fits}
\end{figure}

Figure \ref{fits} shows the low temperature power law fits ($\rho - \rho_0 = AT^n$) to the $\rho(P,T)$ data of \unial.  A  linear $T$-dependence describes the data for $P \leq 25.6$ kbar, whereas an  exponent $n=1.3-1.7$ is needed to fit the $\rho(T)$ curves between P=31.3 kbar and 42.4 kbar over nearly a decade in temperature below 10 K.  At higher pressures of $P=46.4$ kbar and 51.3 kbar, a Fermi-liquid ground state is revealed ($n=2$); however, a NFL-like exponent $n=1.5$ is again found at the highest pressure $P=55.2$ kbar where the N{\'{e}}el temperature ($T_N=2.4$ K) is almost completely suppressed to $T=0$ K,  possibly  indicating the influence of critical spin fluctuations at a QCP of $P_c \sim 60$ kbar.  

\begin{figure}[htbp]
\includegraphics[width=6in]{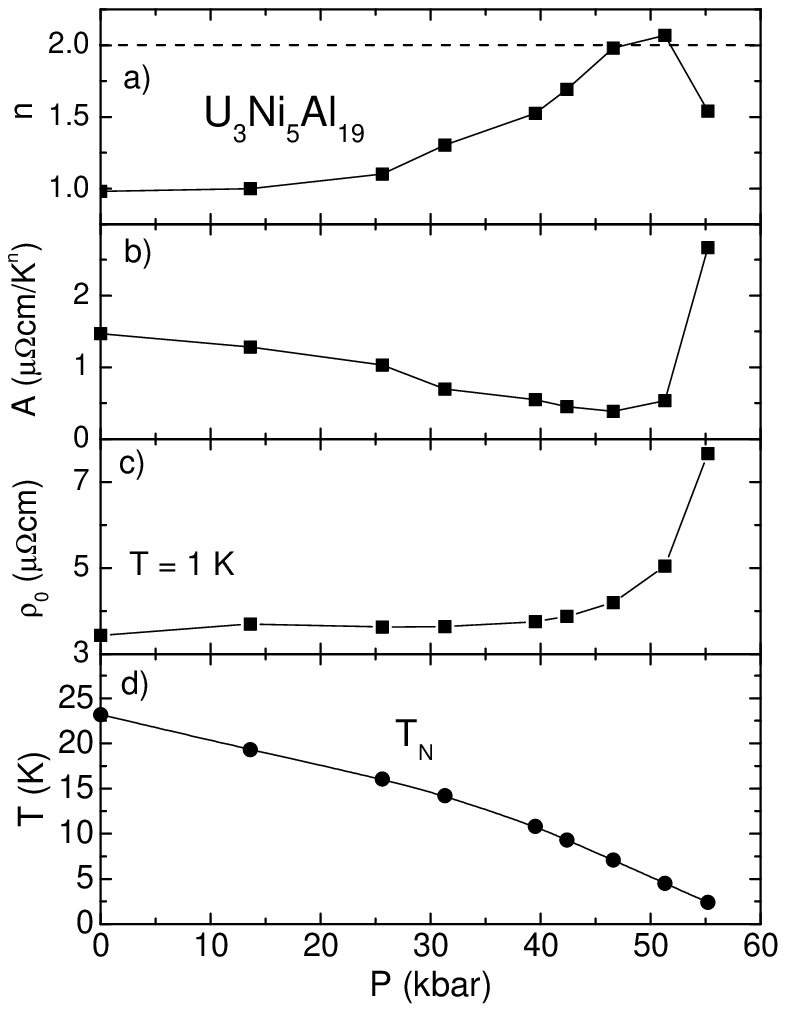}  
 \caption{ Power law exponent $n$, coefficient $A$, residual resistivity $\rho_0$, and N{\'{e}}el temperature $T_N$ vs $P$ in a) - d), respectively.}
\label{props}
\end{figure}

Figure \ref{props} provides a summary of our electrical resistivity measurements under pressure on \unial.  The power law $T$-dependence of $\rho(T)$ for $P \leq 42.4$ kbar indicates a NFL ground state in the presence of long-range magnetic order.  A Fermi-liquid ground state is stabilized at intermediate pressures between 42.4 kbar and 51.3 kbar; at the highest pressure $P=55.2$ kbar, critical fluctuations associated with the assumed AFM QCP at $P_c \sim 60$ kbar once again lead to a NFL exponent of the electrical resistivity of $n=1.5$ (Fig. \ref{props}a).  Both the $A$ coefficient obtained from the power law fits and the residual resistivity (at $T=1$ K) increase dramatically upon approaching the AFM QCP (Fig. \ref{props}); such critical scattering arising from proximity to a QCP is typical of many heavy fermion systems such as CeAgSb$_2$,\cite{Sidorov03} CePd$_2$Si$_2$,\cite{Mathur98} and CeIn$_3$.\cite{Knebel02}

Our results on \unial{} provide evidence for both a quantum critical point at ambient pressure and at $P_c \simeq 60$ kbar.   The application of pressure drives the system away from the $P=0$ QCP, giving rise to FL behavior between $P=42.2-51.3$ kbar, and at the same time suppresses the magnetic transition.  A second quantum critical point at $P_c \sim 60$ kbar presumably leads to the anomalous behavior of $\rho_0$, $A$, and $n$ above 51.3 kbar. Similar behavior has been observed in the heavy-fermion antiferromagnet \ceni, which has three inequivalent Ce sites.  In this compound, AFM order is suppressed by modest pressures of $P_c \simeq 4$ kbar.  Non-Fermi liquid behavior [i.e., $C/T \sim -ln(T/T_0)$] is  found within the AFM state,\cite{Umeo98} possibly associated with one or more of the Ce sites.  The specific heat and electrical resistivity of \ceni{} at $P=4.3$ kbar reveal an evolution to Fermi-liquid behavior above the critical pressure $P_c$;\cite{Umeo98,Umeo96b} however, the magnetic susceptibility still exhibits NFL power law behavior at this pressure.  In \unial, the spin-fluctuation theory of Moriya-Takimoto describes the specific heat at ambient pressure reasonably well; but this model predicts a $T^{3/2}$ variation of the electrical resistivity that is not observed experimentally.  In contrast, the behavior such as the linear variation of $T_N$ with control parameter $(P_c -P)$, the $T^{3/2}$-dependence of $\rho(T)$ near $P_c$, and the steep increase of $A$ close to $P_c$ are consistent with the predictions for a two-dimensional AFM quantum critical point.\cite{Millis93}  

An important issue that remains unresolved is whether  one or more of the distinct U sites is involved in the quantum criticality and antiferromagnetic order at ambient pressure in \unial.  Three possibilities exist: 1) one U site is responsible for both the NFL behavior and the AFM order;   2) the NFL behavior involves one U site while antiferromagnetism is associated with the remaining U site; or 3) both U sites are responsible for  the NFL behavior and the AFM order.  Unfortunately, we can conclude little from the current measurements.  Scenarios 1) and 3) are at least plausible as the coexistence of magnetism and NFL characteristics has been observed before, both in antiferromagnets\cite{note1} and ferromagnets.\cite{Bauer04c} The reduced electronic entropy  in \unial{} tends to favor scenario 1) in which the singly occupied U site is associated with both NFL and AFM phenomena and the remaining doubly occupied U site exhibits temperature independent paramagnetism.  It is interesting to note that the reduced electronic entropy in \unial{} [$S_{el}(T_N) \sim 3.9$ mJ/mol-U K] is similar to U-based antiferromagnets such as U$_2$Zn$_{17}$ [$S_{el}(T_N) \sim 3.8$ mJ/mol-U K] and UCd$_{11}$ [$S_{el}(T_N) \sim 6.5$ mJ/mol K], although a sizable fraction of R$ln(9)$ is released below $T_N$ in other compounds (e.g., UCu$_5$, UAgCu$_4$).\cite{Ott87}   However, model 2) provides a natural explanation for the occurrence of AFM order and NFL behavior at ambient pressure and has been suggested previously.\cite{Haga04}  Neutron scattering measurements on \unial{} would be invaluable for determining which U site(s) is (are) responsible for the magnetic ordering.

In summary, measurements of magnetic susceptibility, specific heat, and electrical resistivity at applied pressures up to 55 kbar have been carried out on single crystals of the  heavy-fermion antiferromagnet \unial.  The logarithmic $T$-dependence of the specific heat and $T-$linear electrical resistivity below 5 K indicates non-Fermi liquid behavior at ambient pressure in the presence of  bulk antiferromagnetic order at $T_N=23$ K.  Electrical resistivity measurements reveal a crossover from NFL to FL behavior at intermediate pressures between 46 kbar and 51 kbar.  The pressure dependence of the physical properties suggest proximity to an AFM QCP at a critical pressure $P_c \approx 60$ kbar.

We would like to thank Z. Fisk for helpful discussions.  Work at Los Alamos was performed under the auspices of the U.S. DOE.
V.A.S. acknowledges the support of Russian Foundation for Basic Research
Grant No. 03-02-17119 and Programs "Strongly Correlated Electrons" and "High
Pressures" of the Department of Physical Sciences, Russian Academy of
Sciences. E. D. B. acknowledges from the G. T. Seaborg Institute for financial support.


\end{document}